\newif\iftwo % twocolumn
\global\twotrue

\iftwo
\documentclass[twocolumn]{aastex631}
\else
\documentclass[manuscript,linenumbers]{aastex631}
\fi

\def\bls{0.98} % 0.98 1.0
\renewcommand{\baselinestretch}{\bls}

\def\ver{1.0}
\newcommand{\orb}{\mbox{\tt orbitN}}
\newcommand{\orbv}{\mbox{\tt orbitN-\ver}}
\newcommand{\hnb}{\mbox{\tt HNBody}}
\newcommand{\reb}{\mbox{\tt REBOUND}}
\newcommand{\rebx}{\mbox{\tt REBOUNDx}}

\newcommand{\grpot}{\mbox{\tt gr\_potential}}
\newcommand{\gr}{\mbox{\tt gr}}

\newcommand{\jtwo}{\mbox{\tt J2}}
\newcommand{\lun}{\mbox{\tt LUNAR}}
\newcommand{\luns}{\mbox{\tt LUN}}
\newcommand{\pn}{\mbox{\tt PN}}

\newcommand{\tcr}{\textcolor{black}}

\newcommand{\tck}{\textcolor{black}}

\renewcommand{\v}[1]{\mbox{\boldmath$#1$}}
\newcommand{\x}{\times}
\newcommand{\e}[1]{\mbox{$\x10^{#1}$}}

\newcommand{\beqn}{\begin{eqnarray}}
\newcommand{\eeqn}{\end{eqnarray}}
\newcommand{\q}{\frac}
\newcommand{\dt}{\Delta t}
\newcommand{\dr}{\partial}
\newcommand{\sm}{\mbox{$\sim$}}
\newcommand{\qd}{\quad}
\newcommand{\qq}{\qquad}
\newcommand{\ausds}{\mbox{au$^2$~d$^{-2}$}}
\newcommand{\asy}{\mbox{$''$/y}}

\newcommand{\MN}{\mbox{$M_0$}}
\newcommand{\Hc}{{\cal H}}
\newcommand{\Dc}{{\cal D}}
\newcommand{\Kc}{{\cal K}}

\newcommand{\Yc}{{\mathit{\Gamma}}}
\newcommand{\Cc}{{\cal S}}
\newcommand{\Fc}{{\cal F}}

\newcommand{\alp}{\alpha}
\newcommand{\bet}{\beta}
\newcommand{\gam}{\gamma}
\newcommand{\lam}{\lambda}
\newcommand{\om}{\omega}
\newcommand{\Om}{\Omega}
\newcommand{\vpi}{\varpi}
\newcommand{\D}{\Delta}
\newcommand{\sig}{\sigma}
\newcommand{\tauD}{\widehat{\tau}_D}

\newcommand{\giturl}{\url{github.com/rezeebe/orbitN}}
\newcommand{\myurl}{\url{www2.hawaii.edu/~zeebe/Astro.html}}
\newcommand{\npurl}{\url{www.ncdc.noaa.gov/paleo/study/35174}}

\shorttitle{\orb}
\shortauthors{Zeebe}

\graphicspath{{./}{figures/}}

\def\figdir{}

\begin{document}

\title{orbitN: A symplectic integrator for planetary systems dominated 
by a central mass - Insight into long-term solar system chaos}

%% The \author command is the same as before except it now takes an optional
%% argument which is the 16 digit ORCID. The syntax is:
%% \author[xxxx-xxxx-xxxx-xxxx]{Author Name}

%\correspondingauthor{}
\email{orbitN.code@gmail.com ; zeebe@soest.hawaii.edu}

\author[0000-0003-0806-8387]{Richard E. Zeebe}
\affiliation{
SOEST, University of Hawaii at Manoa, 
1000 Pope Road, MSB 629, Honolulu, HI 96822, USA. \\ \\
{{\rm Published, doi.org/10.3847/1538-3881/acd63b} \\
The Astronomical Journal}
}

\begin{abstract}

Reliable studies of the long-term dynamics of planetary systems require 
numerical integrators that are accurate and fast. 
The challenge is often formidable because the chaotic 
nature of many systems requires relative numerical error bounds at or close to 
machine precision (${\sim}10^{-16}$, double-precision arithmetic),
otherwise numerical chaos may dominate over physical chaos.
Currently, the speed/accuracy demands are usually only
met by symplectic integrators.
For example, the most up-to-date long-term astronomical solutions for the 
solar system in the past (widely used in, e.g., astrochronology 
and high-precision geological dating) have been obtained 
using symplectic integrators. Yet, 
the source codes of these integrators are unavailable.
Here I present the symplectic integrator \orb\
(lean version 1.0) with the primary goal of generating accurate and 
reproducible long-term orbital solutions for near-Keplerian
planetary systems 
(here the solar system) with a dominant mass \MN. Among other features,
\orbv\ includes \MN's quadrupole moment, a lunar contribution, and
post-Newtonian corrections (1PN) due to \MN\ (fast symplectic 
implementation). To reduce numerical
roundoff errors, Kahan compensated summation was implemented.
I use \orb\ to provide
insight into the effect of various processes on the long-term chaos in
the solar system. \tcr{Notably}, 1PN corrections have the opposite effect on
\tcr{chaoticity/}stability on 100-Myr vs.\ Gyr-time scale.
For the current application, \orb\ is about as fast or faster
(factor 1.15-2.6) than comparable integrators, depending on 
hardware. The \orb\ source code~(C) is available at \giturl.

\end{abstract}



\section{Introduction} \label{sec:intro}

Trustworthy long-term dynamical studies of planetary systems 
require accurate and fast numerical integrators.
The requirements for accuracy and speed are usually mutually 
exclusive because numerical algorithms generally have to 
sacrifice accuracy for speed.
The chaotic behavior of many $N$-body systems presents
a particularly daunting challenge and can produce misleading 
results if numerical chaos dominates over physical chaos
\citep[e.g.,][]{wisdom92,rauch99,hernandez22}. 
As a result, the desired error 
tolerance of numerical integrator schemes is at or close 
to machine precision, i.e.,
about $10^{-16}$ at double-precision floating-point arithmetic.
The tool of choice to tackle the problem
is usually symplectic integrators, which 
show favorable performance in terms of speed, as well as
conservation of energy and angular momentum
\citep[e.g.,][]{wisdom91,yoshida90}. One example of highly
demanding $N$-body integrations are up-to-date long-term 
solar system integrations that provide astronomical solutions
for the past and are widely used in, for instance, astrochronology 
and high-precision geological dating 
\citep{laskar11,zeebelourens19,zeebelourens22epsl}.
In addition to accurate and fast integration of the
fundamental dynamical equations for the main solar 
system bodies, generating adequate
astronomical solutions requires proper inclusion of
(1) the Sun's quadrupole moment $J_2$, (2) the effect
of the Moon, (3) post-Newtonian corrections from general 
relativity (1PN), and (4) a contribution from asteroids.
Note that while the effects from $J_2$ and asteroids
may appear negligible, their contributions become critical
for astronomical solutions over, e.g., 50-Myr time scale due 
to chaos. The most recent
astronomical solutions have been obtained using a higher-order
symplectic integrator scheme called SABAC$_4$ 
\citep{laskar11} and a 2nd order symplectic scheme
available in the integrator package \hnb\ \citep{rauch02,zeebe17aj}. 
The executable and source code of the SABA integrators have 
not been made available for researchers to use \citep{laskar11}.
Binaries of the \hnb\ package are available, while the 
source code is unavailable \citep{rauch02}.

Alternative symplectic integrator packages with available 
source code that could potentially be used to generate 
accurate $N$-body/astronomical solutions include 
{\tt swift/swifter}, {\tt mercury6}, and \reb\ 
\citep{levison94,duncan98,chambers99,kaufmann05,rein15}. 
However, {\tt swift/swifter} and 
{\tt mercury6} do not provide features such as fast and 
accurate options to include 1PN corrections,
which are critical for the current application.
While the \reb/\rebx\ package does provide 1PN 
options \citep{reinliu12,rein15,tamayo20}, their accuracy or 
performance turned out to be suboptimal for the current problem 
(see Section~\ref{sec:gr}).
In this contribution, I present the symplectic integrator \orb\
(lean version 1.0) with the primary goal of efficiently
generating accurate and reproducible long-term orbital 
solutions for near-Keplerian planetary systems dominated by a 
central mass. \orb\ version 1.0 focuses on hierarchical systems 
without close encounters but can be extended to include
additional features in future versions.
The \orb\ source code~(C) is available at \giturl\
(correspondence to orbitN.code@gmail.com).
While the current \orb\ application focuses 
on the solar system, \orb\ can generally be applied to planetary systems 
with a dominant mass. The present solar system integrations 
with \orb\ reveal that 1PN corrections have the opposite effect on
\tcr{chaoticity/}stability on 100-Myr vs.\ Gyr-time scale. 

\section{Hamiltonian splitting} \label{sec:ham}

The core of \orb's integrator scheme is based on a 2nd order 
symplectic map, which is described at length elsewhere and
is not repeated here \citep[e.g.,][]{wisdom91,yoshida90,
saha94,mikkola97,chambers99,murraydermott99,rein15}.
[Note that recent studies indicate that higher order symplectic
schemes are not necessarily advantageous \citep{hernandez22,abbot23}].
A few features deserve attention here such as the Hamiltonian 
splitting and mass factors, which are important, for instance, for 
the implementation of 1PN corrections.

The gravitational $N$-body Hamiltonian may be split into a Kepler-
and Interaction part \citep[e.g.,][]{murraydermott99,rein15}:
\beqn
\Hc = \Hc_{Kep} + \Hc_{Int} \ ,
\label{eq:ham}
\eeqn
where
\beqn
\Hc_{Kep} = \sum_{j=1}^J \left(
            \q{p'^2_j}{2m'_j} - \q{\mu_j m'_j}{r'_j}
                         \right)
\eeqn
and \citep{rein15}:
\beqn
\Hc_{Int} = \sum_{j=1}^J
            \q{\mu_j m'_j}{r'_j} 
          - \sum_{j=0}^{J-1} \sum_{k=j+1}^J \q{G m_j m_k}{r_{jk}} \ ,
\label{eq:int}
\eeqn
where $J = N-1$ and $N$ is the total number of bodies in the 
system, including the dominant mass (index 0); $p$ refers
to momentum, $m$ to mass, $r$ to distance, e.g., 
$r_{jk} = |\v{x}_k - \v{x}_j|$ and $G$ is
the gravitational constant in appropriate units. Primes 
indicate quantities in Jacobi coordinates {\citep[for 
a summary of coordinate choices and operator splitting, 
see][]{hernandez17}. The factor
$\mu_j$ is given by $\mu_j = G \cdot \sig_j$, where
\beqn
\sig_j = \sum_{i=0}^j m_i \ .
\eeqn
Importantly, the mass factor in the first term of 
Eq.~(\ref{eq:int}) depends on the Hamiltonian splitting,
which differs, for instance, between \citet{rein15} and
\citet{saha94} (ST94 hereafter).
While the Hamiltonian splitting in \orb\ 
follows \citet{rein15}, the 1PN corrections in \orb\ 
follow ST94. Hence the $\mu_j$ factors that enter 
the equation for 1PN corrections here are $\mu_j = G \cdot \sig_j$ 
(see Section~\ref{sec:pn}), and not $\mu_j = G \cdot 
\sig_j/\sig_{j-1}$ as in ST94.

\section{Architecture} \label{sec:arch}

\orb's structure consists of basic function sequences, including
input (masses $m_j$, initial positions $\v{x}^0_j$ and velocities 
$\v{v}^0_j$), integration (operator application such as Drift 
and Kick, see below), and output as
requested. \orb\ is written in C (C99 standard) and intentionally uses 
standard function calls with (usually) explicitly stated arguments 
such as $f(\v{x}_j, \v{v}_j, \ldots)$ to highlight the function's
input- and output/updated variables. Large data structures ({\tt struct} 
in C/C++), which by design often hide the input- and output/updated 
variables, are avoided. Different integration sequences are available 
in \orb, including {\tt slow}, {\tt fast}, and {\tt fast\_pn}, as
explained below.

The core of \orb's integrator scheme is based on a 2nd order 
symplectic map, frequently referred to as a Wisdom-Holman
(WH) map \citep{wisdom91}.
The time evolution under the Hamiltonian split (Kepler and 
an Interaction part, see Eq.~(\ref{eq:ham})) is realized by Drift 
and Kick operators, $\Dc(\tau)$ and $\Kc(\tau)$ (generally
functions of $\v{x}_j, \v{v}_j, \ldots$), where the operator 
timestep argument 
$\tau$ is usually a simple function of the fixed integration 
timestep $\dt$. \orb's 2nd order integrator is based on a 
Drift-Kick-Drift operator sequence, advancing the state 
variables $\v{x}_j$ and  $\v{v}_j$ (e.g., 
Appendix~\ref{sec:alpha}, Eq.~(\ref{eq:zdot})).
The operator code sequence {\tt slow} reads (timestep counter 
$i = 1, \ldots, n_{step}$):
\beqn
\Dc (\dt/2) \circ \Kc (\dt) \circ  \Dc (\dt/2) 
\ \circ \ \ldots \
\iftwo \nonumber \\ \fi
  \circ \
\Dc (\dt/2) \circ \Kc (\dt) \circ \Dc (\dt/2) \ .
\eeqn
Except for the first and final step, and when output is requested, 
the interior $\dt/2$-Drift steps can be combined into
a single step:
\beqn
\Dc (\dt/2) \quad \circ
\iftwo \hspace*{35ex} \nonumber \\ \fi
\quad \Kc (\dt) \circ  \Dc (\dt)
\ \circ \ \ldots \  \circ \
\Dc (\dt) \circ \Kc (\dt) \quad \circ
\iftwo \nonumber \\ \fi
\quad \Dc (\dt/2) \ ,
\iftwo \hspace*{31.5ex} \fi
\eeqn
representing the operator code sequence {\tt fast}. When
including 1PN corrections, additional operators are applied
(option {\tt fast\_pn}, see Section~\ref{sec:pn}). Ignoring
initial and final steps, and including Kahan compensated 
summation (operator $\Cc$), the {\tt fast\_pn} core sequence, 
for example, reads:
\beqn
\ldots \           
                  \circ \
\Yc\Dc\Yc(\tau)   \circ
\Cc               \circ
\Kc (\dt)         \circ
\Cc               \circ
\Yc\Dc\Yc(\tau) \ \circ \ 
\ldots \ ,
\eeqn
where 
\beqn
\Yc\Dc\Yc(\tau) = 
\Yc (\dt/2) \circ
\Cc         \circ
\Dc (\dt)   \circ
\Cc         \circ
\Yc (\dt/2)
\eeqn
and $\Cc = \Cc(\v{x}_j,\v{v}_j, \ldots)$ (for different Kahan 
summation options, see Section~\ref{sec:khn}) and $\Yc(\tau)$ 
represents the $\gam$-term of the 1PN Hamiltonian 
(see Section~\ref{sec:pn}).

\orb\ version 1.0 uses Gaussian units, i.e., length, time, and 
mass are expressed in units of au, days, and fractions of \MN, although this
feature can be extended to other sets of units in future versions.
Orbital coordinates in \orb\ can be output as state vectors $\v{x}_j, 
\v{v}_j$ or Keplerian elements. However, for accuracy and archiving 
of results, state vectors are recommended (see Section~\ref{sec:trig}).
\orb's source code is provided to the user and compiled on the local
machine. \orb\ has been tested on linux and mac platforms. For example,
a full solar system integration over 100 Myr, comprising the planets, 
Pluto, and ten asteroids, and including \MN's quadrupole moment, a lunar 
contribution, and 1PN corrections requires about 16h wall-clock time 
on a 64-bit Linux machine ({\tt gcc} optimization level 3, Intel i5-10600 
@3.30GHz, see also Section~\ref{sec:gr}). 

\subsection{Trigonometric functions} \label{sec:trig}

Trigonometric functions are frequently employed in numerical 
integrators, including the WH map. For example, the 
Drift operator may use Gauss' classic $f$ and $g$ functions 
to advance $(\v{x}_j, \v{v}_j)$ (see Section~\ref{sec:khn}),
which includes numerical sine and cosine evaluations of the 
eccentric anomaly \citep[e.g.,][]{danby88}. The problem with
numerical evaluations of trigonometric functions is that 
different compilers and architectures may produce different
results for the same operation. The 2019 IEEE-754 Standard for 
Floating-Point Arithmetic handles trigonometric functions
under ``Recommended Operations'', which are not mandatory 
requirements \citep{ieee19}. As a result, even if 
floating-point operations on a given platform adhere to 
the IEEE-754 standard, there is no guarantee that the
results of trigonometric operations are identical between
different platforms. For example, I tested evaluation of 
Gauss' $f$ and $g$ functions on various linux machines 
(including the same binary) with the same operating system 
but different 
hardware, which yielded different results. Although initially 
close to machine precision, the differences can grow very
quickly, e.g., for chaotic systems, which renders the results
of such integrations practically irreproducible
\citep{zeebe15apjA,zeebe15apjB,zeebe22aj}.
The problem extends to different architectures, compilers, 
optimization levels, etc. 

\citet{itokojima05} discussed the unsatisfactory status
of trigonometric functions in computer mathematical 
libraries and reported their communications with computer 
manufacturers about the issue. As a workaround, 
\citet{itokojima05} suggested optimizing and porting 
certain mathematical libraries. I unsuccessfully tested several 
alternative methods, including compiling sine 
and cosine functions directly from source code and a 
discretization/Taylor expansion approach using lookup tables 
of trigonometric functions \citep{fukushima97}. Most
methods turned out to be cumbersome to implement and 
showed poor performance. A satisfactory alternative that
also showed good performance is based on Stumpff functions 
\citep{stumpff59}, which avoids evaluation of trigonometric 
functions altogether (see Section~\ref{sec:stumpff}).

Note in this context that because the conversion of orbital 
coordinates from state vectors $\v{x}_j, \v{v}_j$ to Keplerian 
elements involves trigonometric functions, state vectors are 
recommended for output in \orb, for instance, when accuracy is 
required and for archiving of results (see above).
\tck{If needed, a separate routine is provided
for post-run conversion of state vectors to Keplerian 
elements. In that case, and if the selected
conversion involves the masses of individual bodies, the user
is required to provide the original mass/coordinate 
input file of the run.}

\subsection{Stumpff- and $f$ and $g$ functions \label{sec:stumpff}}

The Drift operator (also often called Kepler solver) can be 
formulated using universal variables and Stumpff functions,
the particulars of which have been described elsewhere and are not 
repeated here \citep{stumpff59,stiefel71,danby87,danby88,mikkola97,
mikkolaarseth98,mikkolainnanen99,rein15}. In the following,
a few details are noted that pertain to the implementation of
Stumpff functions in \orb, which follows \citet{rein15}.
The $f$ and $g$ functions used with universal variables and 
Stumpff functions to advance the Kepler drift from time $t$ to 
$t + \tau$ may be written as:
\beqn
     f  =     1 - \mu G_2 / r         \qq & ; & \qq 
     g  =  \tau - \mu G_3             \label{eq:fg}    \\
\dot{f} = -\mu G_1 / (r \cdot r_\tau) \qq & ; & \qq 
\dot{g} =     1 - \mu G_2 /  r_\tau   \label{eq:fgdot} \ ,
\eeqn
where $\mu = \mu_j$ (see Section~\ref{sec:ham}),
$r_\tau = r(t+\tau) = r + \chi G_1 + \zeta G_2$, 
$\chi = \v{x} \cdot \v{v}$,
$\zeta = \mu - \bet r$, and $\bet = 2\mu/r - v^2$.
The so-called $G$-functions are given by:
\beqn
G_n(\bet,s) = s^n \cdot c_n(x) \qq ; \qq x = \bet \ s^2 \ ,
\eeqn
where $s$ is the variable solved for in Kepler's equation in universal 
variables \citep{danby87,danby88} and the $c_n(x)$ are Stumpff 
functions, or $c$-functions \citep[][Eq.~(V; 43)]{stumpff59}:
\beqn
c_n(x) = \q{1}{n!} - \q{x}{(n+2)!} + \q{x^2}{(n+4)!} - \ldots
\qq ; \qq 
\iftwo \nonumber \\ \fi
n = 0,1,2,\ldots \ .  \qquad
\eeqn
Note that Gauss' $f$ and $g$ functions and those given in 
Eqs.~(\ref{eq:fg}) and~(\ref{eq:fgdot}) have different 
individual terms but the overall structure is the same
(see Section~\ref{sec:khn}). For the problems studied here,
an initial guess $s_0$ for $s$ based on Eq.~(18) in 
\citet{danby87} worked well in \orb:
\beqn
s_0 = \q{1}{r} \ \tau - \q{\chi}{2r^3} \ \tau^2 \ .
\eeqn

Because the Stumpff functions are based on a series expansion,
the series may be truncated at lower $n$ for small 
$x$, given a required accuracy (say machine
precision). For planetary systems with a large range in $x$, the
integrator performance can thus be improved by reducing $n$ 
depending on $x$; a simple option to do so is
available in \orb. The primary choice to solve Kepler's
equation is usually based on the Newton-Raphson method
(fast and generally accurate), which is also the case in \orb. 
As a secondary choice (in case Newton-Raphson fails), the 
secant method \citep[e.g.,][]{danby88} was implemented in \orb. 
As a third and final choice, the bisection method is 
used \citep{rein15}.

\subsection{Kahan compensated summation \label{sec:khn}}

To reduce numerical roundoff errors, Kahan compensated 
summation (operator $\Cc$, see Section~\ref{sec:arch}) 
was implemented in \orb\ 
\citep{kahan65}. At least two different implementation 
options are possible. (1) $\Cc$ is applied after each operation 
that updates the state variables, say $[\v{x}_j \ \v{v}_j]_{t+\tau} = 
[\v{x}_j \ \v{v}_j]_t + [\D \v{x}_j \ \D \v{v}_j]$, i.e., 
multiple times per timestep $\dt$. (2) $\Cc$ is applied
only once per timestep and the incremental updates from 
different operators during $\dt$ are accumulated into a 
separate set of variables $(\delta \v{x}_j,\delta \v{v}_j)$,
representing changes in state variables, to which $\Cc$
is applied. The advantage of option
(1) is that carrying $(\delta \v{x}_j,\delta \v{v}_j)$ through
the integration is avoided; its disadvantage is that $\Cc$
has to be applied multiple times per timestep. The advantage of
option
(2) is that $\Cc$ has to be applied only once per timestep; its
disadvantage is that $(\delta \v{x}_j,\delta \v{v}_j)$ has to 
be carried through the integration. Furthermore,
internally, the updated sum $(\v{x}_j + \delta \v{x}_j)$
and/or $(\v{v}_j + \delta \v{v}_j)$ has to be evaluated during 
the timestep regardless because its value is required
as input to a subsequent operator. I tested both 
options and found no significant differences in results or 
performance. Option~(1) was implemented in \orbv, which
adds a computational overhead of about~3\% for a full
solar system integration.

Kahan compensated summation generally uses increments, 
say $(\D \v{x}, \D \v{v})$, and hence is straightforward 
to apply following the Kick operator
because accelerations, $\v{a}$, and $\D \v{v}
= \v{a} \x \tau$ are explicitly calculated in the Kick
routine. However, the Drift operator uses $f$ and $g$ 
functions (Eqs.~(\ref{eq:fg}) and~(\ref{eq:fgdot})) to advance
$(\v{x}, \v{v})$ for each body from time $t$ to $t + \tau$,
which is not expressed in terms of increments:
\beqn
\v{x}(t+\tau) & = &       f \cdot \v{x}(t) 
                  +       g \cdot \v{v}(t) \label{eq:xtau} \\
\v{v}(t+\tau) & = & \dot{f} \cdot \v{x}(t) 
                  + \dot{g} \cdot \v{v}(t) \label{eq:vtau} \ .
\eeqn
As mentioned above,
the $f$ and $g$ functions in \orb\ are adjusted for use with 
universal variables and Stumpff functions and differ from 
Gauss' $f$ and $g$ functions, although they have the same 
structure (see Section~\ref{sec:stumpff}).
Inserting Eqs.~(\ref{eq:fg}) and~(\ref{eq:fgdot}) into
Eqs.~(\ref{eq:xtau}) and~(\ref{eq:vtau}), the last two
equations may be rewritten in terms of $\D \v{x}$ and
$\D \v{v}$:
\beqn
\D \v{x} & = &      \widehat{f}  \cdot \v{x}(t) 
             +               g   \cdot \v{v}(t) \\
\D \v{v} & = &          \dot{f}  \cdot \v{x}(t) 
             + \widehat{\dot{g}} \cdot \v{v}(t) \ ,
\eeqn
where $\widehat{f} = f -1$ and $\widehat{\dot{g}} = \dot{g} -1$
\citep[see also][who used an analogous procedure with Gauss' $f$ and 
$g$ functions]{wisdom18}.

\subsection{Symplectic correctors}

Symplectic correctors remove fluctuations in energy and
were implemented up to stage~6 = 7th order, following
\citet{wisdom06}. Stages~2, 4, and~6 are available in \orbv\
and are applied only at the beginning and end of an 
integration, and when output is requested --- hence adding
essentially no computational overhead.
Plots of relative maximum energy changes indeed suggest
reductions by up to two orders of magnitude when including 
the corrector (stage~6). Interestingly, however, 
for practical applications over, say, the past 60~Myr 
or so, symplectic correctors make little difference as the 
actual dynamics in terms of orbital eccentricity, mean longitude,
etc. are hardly affected over that time scale
(see Section~\ref{sec:std}).

\subsection{\MN's quadrupole moment}

The gravitational quadrupole potential due to the dominant/central
mass \MN\ may be written as:
\beqn
\Phi_{J2} & = & \q{G \MN}{r} \ J_2 \ (R_0/r)^2  \
                \q{1}{2} (3 \cos^2 \theta - 1) \\
          & = & \q{A}{r^3}  \left( \q{3z^2}{r^2} - 1 \right) \ ,
\eeqn
with $A = G \MN \ J_2 \ R_0^2 /2$ and $z = r \cos \theta$
(see below for $z$'s reference frame),
where $J_2$ is \MN's quadrupole moment, $R_0$ its effective
radius (related to oblateness), and $\theta$ is the colatitude 
angle. Applying the gradient $-\nabla \Phi_{J2}$, the accelerations
are given by:
\beqn
a_x   =   3A
\left( \q{5z^2}{r^2} - 1 \right) \q{x}{r^5} \qd ; \qd           
a_y   =   3A
\left( \q{5z^2}{r^2} - 1 \right) \q{y}{r^5} \qd ; 
\iftwo \nonumber \\ \fi
\qd a_z   =   3A
\left( \q{5z^2}{r^2} - 3 \right) \q{z}{r^5} \ ,
\iftwo \hspace*{25ex} \fi
\eeqn
as implemented in \orb\ (macro option \jtwo).
Importantly, in case of the solar system, the solar 
quadrupole moment is directed along the solar rotation/symmetry 
axis, which is about $6^\circ$ and $7^\circ$ offset from the invariable 
plane and ECLIPJ2000, respectively.
By default, the quadrupole axis in integration coordinates is 
directed along the $z$-axis. Thus, if the initial 
(Cartesian) coordinates (say, obtained from ephemerides)
are specified in a different frame, then the
coordinates need to be rotated to account for the offset
between that frame and the solar rotation axis \citep{zeebe17aj}.

\subsection{Lunar contribution}

For solar system integrations, the Moon may be included as 
a separate object. Alternatively, the Earth-Moon system may be 
modeled as a point mass at the Earth-Moon barycenter (EMB), plus
an additional effect from the Moon's influence on the net 
EMB motion via a mean quadrupole potential (\orb\ macro option 
\lun). The quadrupole acceleration term may be written as 
\citep{quinn91}:
\beqn
\v{a}_Q = - \q{3 \ G \MN \ m_E \ m_L \ R^2}{4 \ (m_E + m_L)^2} 
            \cdot \q{\v{r}}{r^5} \cdot f_L \ ,
\label{eq:aql}
\eeqn
where indices `$E$' and `$L$' signify Earth and Lunar, $R$ is 
an effective parameter for the Earth-Moon distance,
and $r$ is the EMB-Sun distance. The factor $f_L$ is
a correction factor that deserves a few comments.
\citet{quinn91} introduced $f_L$ to account for differences
between the actual lunar orbit and their simplified
model and set $f_L = 0.9473$. \citet{varadi03}
revisited the issue and suggested $f_L = 0.8525$,
based on a comparison to integrations that resolved
the Moon as a separate object. Using $f_L = 0.8525$,
\citet{zeebe17aj} showed that integrations with a
separate Moon (Bulirsch-Stoer algorithm) and Quinn
et al.'s lunar model (symplectic map) virtually
agreed to \sm63~Myr in the past (i.e., divergence
time $\tauD \simeq 63$~Myr, see Section~\ref{sec:chaos}).
This time scale is 
beyond the solar system's intrinsic predictability limit 
of \sm50~Myr due to dynamical chaos and hence the lunar
model is likely sufficient for most applications
that do not need to resolve the Moon. The \lun\ option was 
implemented in \orb\ following \citet{quinn91,rauch02}.

\subsection{Post-Newtonian corrections from general relativity 
            \label{sec:pn}} 

\orbv\ includes post-Newtonian corrections from general relativity
due to \MN\
(to 1PN order), implemented following ST94. The 1PN
Hamiltonian may be written as \citep{landau71fields,saha94,will14}:
\beqn
\Hc_{PN} & = & \q{1}{c^2} \ \sum_{j=1}^J \left( 
                          \q{\mu^2_j \ m'_j}{2 \ r'^2_j}
                        - \q{p'^4_j}{8 \ m'^3_j}
                        - \q{3 \ \mu_j \ p'^2_j}{2 \ m'_j \ r'_j}
                               \right)
\label{eq:hpn}
\eeqn
where $c$ is the speed of light, $\mu_j = G \cdot \sig_j$ (see 
Section~\ref{sec:ham}), $m$ refers to mass, $r$ to distance, 
and $p$ to momentum. Primes indicate quantities in Jacobi 
coordinates. 
 
Note that at the present level of approximation (1PN), 
the difference between, e.g., Jacobi and bodycentric distances
and masses can be ignored.
Consider the magnitude of the primary
Newtonian potential in Gaussian units at 1~au, $G \MN/r = 
k^2 \MN/r \simeq 10^{-4}$~\ausds, vs.\ the 1st term of the 1PN
potential (Eq.~(\ref{eq:hpn})), $k^4 \MN^2/(2c^2r^2) \simeq 
10^{-12}$~\ausds, i.e., a relative 1PN magnitude of about
$10^{-8}$ (factor $k^2 \MN/(2c^2r)$).
The relative differences between Jacobi- and bodycentric distances,
$\D r/r$ (and masses $\D m/m$), for example, for Mercury and 
Earth are 0 and $\sm 10^{-7}$ ($\sm 10^{-7}$ and $\sm 10^{-6}$).
Such differences would introduce a relative error of 
$\lesssim 10^{-14}$ with respect to the primary 
Newtonian potential, far beyond the level of 
accuracy provided by the 1PN approximation.

The 1PN Hamiltonian (Eq.~(\ref{eq:hpn})) includes cross-terms
between positions and momenta and, as written, does not split
conveniently into terms similar to Eq.~(\ref{eq:ham}). However,
ST94 devised a method to accommodate $\Hc_{PN}$ in a symplectic 
scheme, which also shows favorable performance in terms of energy, 
angular momentum, and speed (see Section~\ref{sec:gr}). 
Eq.~(\ref{eq:hpn}) may be rewritten as:
\beqn
\Hc_{PN} = \sum_{j=1}^J \left(
           \alp_j \ \Hc^2_{Kep,j} + \bet_j / r'^2_j + \gam_j \ p'^4_j
                        \right) \ ,
\label{eq:hpnII}
\eeqn
where $\alp_j = 3/(2 m'_j c^2)$, $\bet_j = -\mu^2_j m'_j /c^2$,
and  $\gam_j = -1/(2 m'^3_j c^2)$. The $\alp$-term merely leads
to a scaling of the timestep argument of the Drift operator 
(see Appendix~\ref{sec:alpha}),
the $\bet$-term can be easily accommodated in the Kick operator, 
and the $\gam$-term can be included as leapfrog operators pre- 
and post Drift step. For the current solar system integrations,
the 1PN option as implemented in \orb\ adds less than \sm10\% 
computational overhead.

\section{Solar System Chaos \label{sec:chaos}}

In the following, I present solar system integrations with \orb\
and other integrator packages to provide insight into the 
integrator algorithms and solar system chaos.
As a chaos indicator, the difference 
between two orbital solutions $i$ and $j$ may be tracked using 
the divergence time $\tauD$ \citep[see][]{zeebe17aj}, i.e., the 
time interval ($\tauD > 0$) after which the maximum absolute difference 
in Earth's orbital eccentricity ($\max | e_i-e_j|$) irreversibly 
crosses \sm10\% of mean $e$ ($\sm0.028\x0.1$, Fig.~\ref{fig:std}). The 
divergence time $\tauD$ as employed here should not be confused with 
the Lyapunov time, which is the time scale of exponential divergence
of trajectories and is only \sm5 Myr for the inner planets.
For the solutions discussed here, the divergence
of trajectories is ultimately dominated by exponential 
growth, which is indicative of chaotic behavior
($t \gtrsim 40$~Myr for standard solar system integrations,
see Section~\ref{sec:std}).
Thus, $\tauD$ is largely 
controlled by the Lyapunov time, although the two are different 
quantities. Integration errors usually 
grow polynomially and typically dominate for $t \lesssim 
40$~Myr (see e.g., Fig.~\ref{fig:std} 
and \citet{varadi03}).

\subsection{Standard test: Past 100~Myr \label{sec:std}}

\begin{figure*}[t]
\begin{center}
\vspace*{-09ex}
\includegraphics[scale=0.8]{\figdir 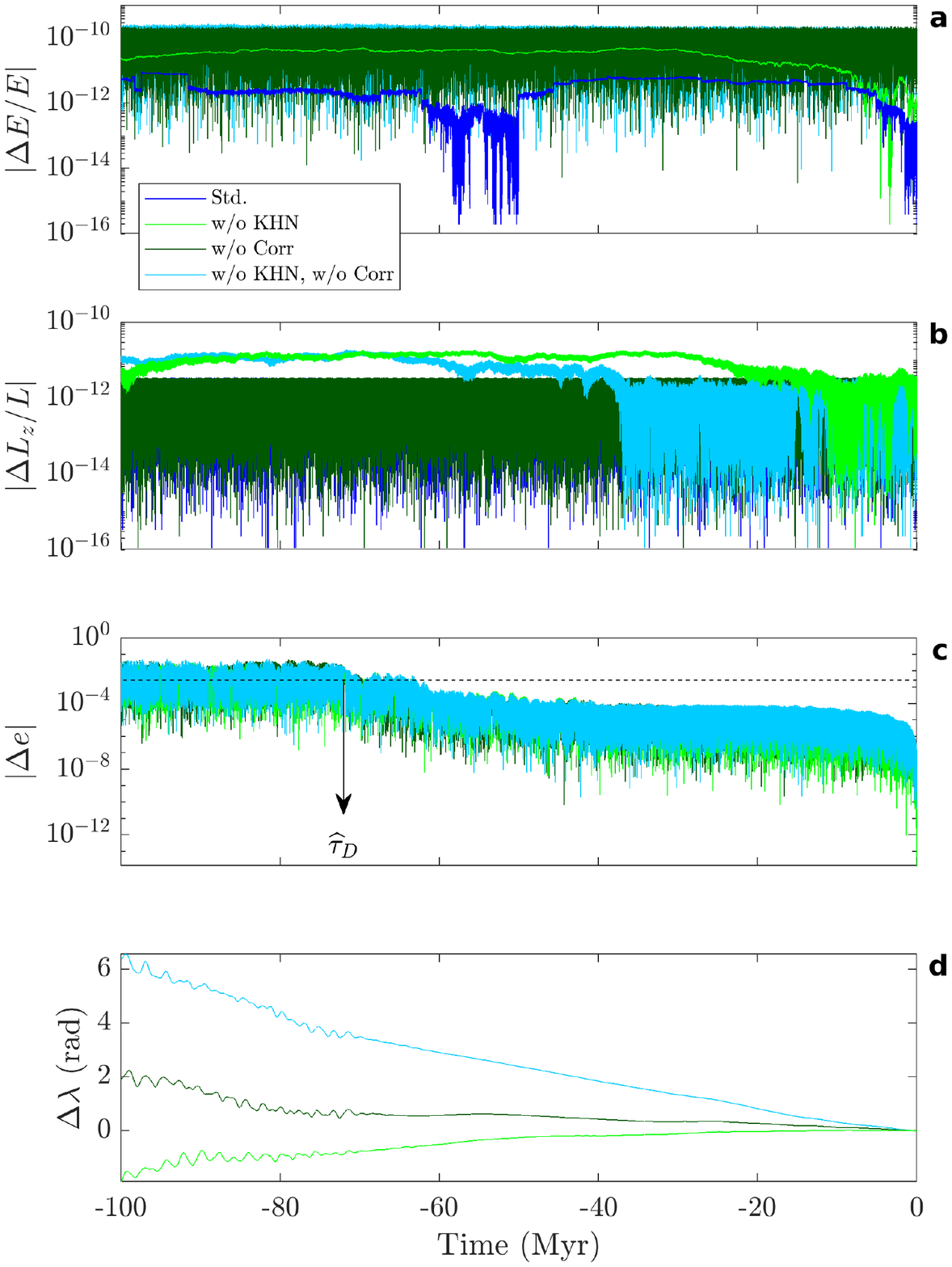}
\end{center}
\vspace*{-16ex}
%\plotone{../eps/x.pdf}
\caption{
Solar system integrations over the past 100~Myr with \orb.
The standard setup (Std.) includes the 
planets, Pluto, ten asteroids, $J_2$, a lunar contribution, and 1PN 
corrections implemented following \citet{saha94}. KHN = Kahan
compensated summation \citep{kahan65}. Corr = symplectic corrector
\citep[stage~6 = 7th order, see][]{wisdom06}. The timestep is $\dt = 
2$~days. $\D E/E = (E-E_0)/E_0$ and $\D L_z/L$ indicate
relative changes in energy and angular momentum.
Differences in orbital eccentricity ($\D e$) and mean 
longitude ($\D \lam$) are for the Earth-Moon barycenter, relative
to the standard run. $\tauD$ is the divergence time (see text).
\label{fig:std}
}
\end{figure*}

For the present standard solar system integrations,
initial conditions for the positions and velocities of 
the planets and Pluto were generated from the JPL DE431
ephemeris \citep{folkner14} 
(\url{naif.jpl.nasa.gov/pub/naif/generic_kernels/spk/planets}),
using the SPICE toolkit for Matlab
(\url{naif.jpl.nasa.gov/naif/toolkit.html}).
We have recently also tested the latest JPL ephemeris DE441
\citep{park21de}, which makes little difference for
practical applications because the divergence time relative to 
the astronomical solution ZB18a (based on DE431) is \sm66~Ma 
(see Section~\ref{ZB18a}) and hence beyond ZB18a's reliability 
limit of \sm58~Ma (based on geologic data, see below).
The standard integration includes 
10 asteroids, with initial conditions generated 
at \url{ssd.jpl.nasa.gov/x/spk.html} (for a list 
of asteroids, see \citet{zeebe17aj}).
Coordinates were obtained at JD2451545.0 
in the ECLIPJ2000 reference frame and subsequently 
rotated to account for the solar quadrupole moment 
($J_2$) alignment with the solar rotation axis 
\citep{zeebe17aj}. Our astronomical solutions are
provided over the time interval from 100-0~Ma. However, 
as only the interval 58-0~Ma 
is constrained by geologic data \citep{zeebelourens19}, we 
caution that the interval prior to 58~Ma is unconstrained due 
to solar system chaos.
The standard integration includes the solar quadrupole moment
$J_2 = 1.305\e{-7}$, the lunar contribution, and 1PN 
corrections. Unless stated otherwise, the integration
timestep is $\dt = 2$~days, as
previously used in our astronomical solution ZB18a 
(see Section~\ref{ZB18a}), which also properly resolves Mercury's
pericenter (at eccentricity $\lesssim 0.2$), hence avoiding 
numerical chaos \citep[see][]{wisdom15,hernandez22}.

As a first test, the long-term behavior of
changes in energy ($\D E/E = (E-E_0)/E_0$), angular momentum 
($\D L_z/L$), and the implementation of Kahan compensated 
summation and symplectic correctors in \orb\ is examined 
(Fig.~\ref{fig:std}). As should be expected from a symplectic
algorithm, $|\D E/E|$ remains small ($|\D L_z/L|$ as well)
and do not
exhibit any significant trends over 100~Myr. Omitting
Kahan summation, maximum $|\D E/E|$ and $|\D L_z/L|$ increase by
up to a factor of \sm10. Omitting the symplectic corrector
\citep[stage~6 = 7th order, see][]{wisdom06}, the energy
fluctuations that are removed by the corrector become 
apparent; the corrector has little effect on $|\D L_z/L|$.
Omitting both Kahan summation and the corrector yields
similar results to omitting just the corrector. The 
consequences for the dynamics of the system may be illustrated 
by examining the differences in the EMB's orbital eccentricity 
($\D e$) and mean longitude ($\D \lam$), relative to the 
standard run (Fig.~\ref{fig:std}c and~d). The effects of 
Kahan summation and the corrector on $\D e$ are similar
to those of a small perturbation or a small difference in 
initial conditions, which grows over time
\citep[see below and][]{zeebe15apjA,zeebe15apjB,zeebe17aj}.
Notably, for practical applications in astrochronology over,
say, the past 60~Myr or so, Kahan summation and symplectic correctors 
would actually make little difference because the divergence time 
to the standard solution is $\gtrsim$~70~Ma (Fig.~\ref{fig:std}c), i.e., 
significantly beyond its reliability limit of \sm58~Ma (see above).

\subsection{Astronomical solution ZB18a \label{ZB18a}}

\begin{figure*}
\begin{center}
\vspace*{-08ex}
\includegraphics[scale=0.8]{\figdir 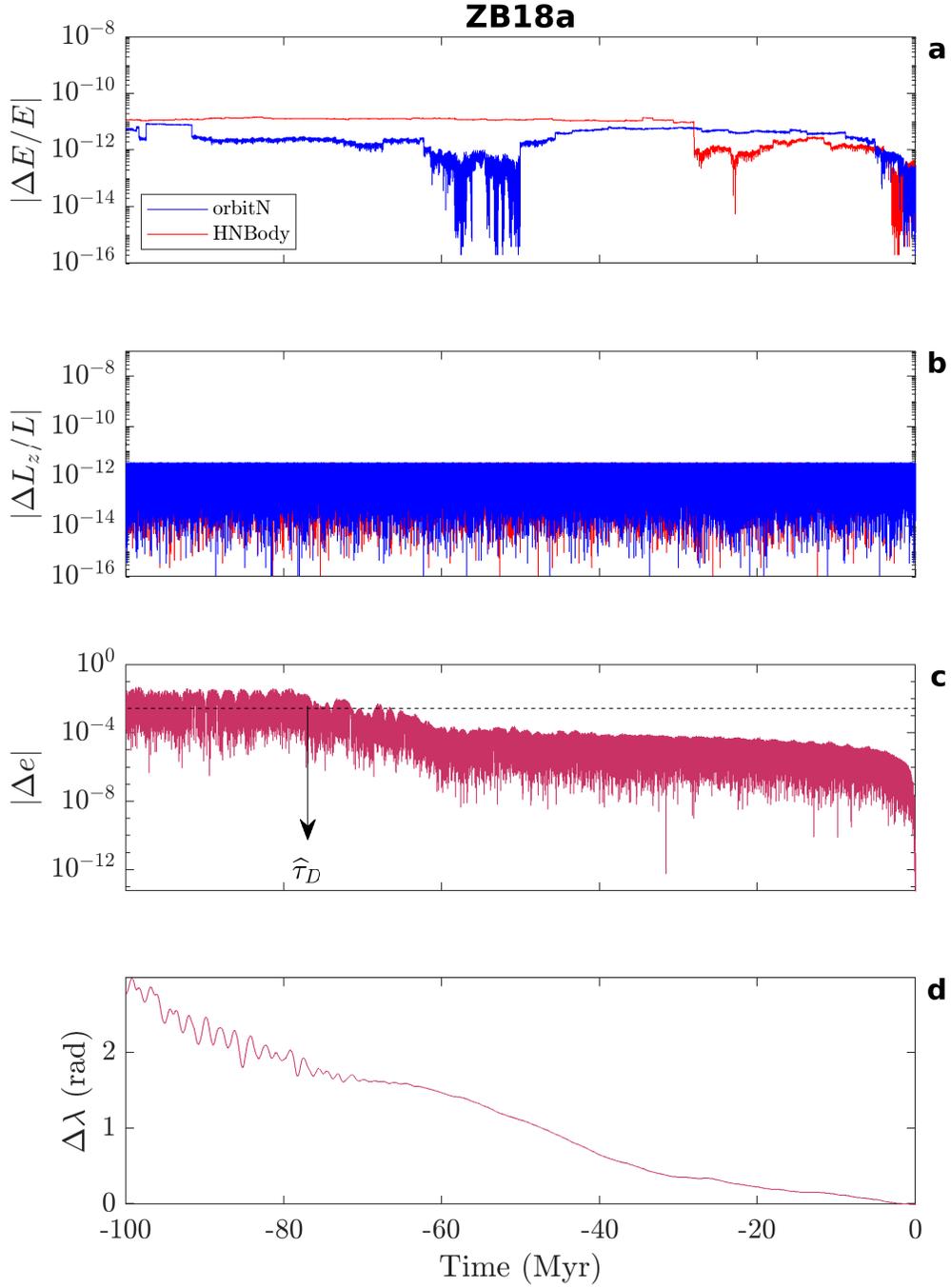}
\end{center}
\vspace*{-16ex}
%\plotone{../eps/x.pdf}
\caption{
Comparison of solar system integrations with \orb\ and
\hnb\ for the setup of the astronomical solution ZB18a 
\citep[][see text for details]{zeebelourens19}.
$\D E/E = (E-E_0)/E_0$ and $\D L_z/L$ indicate
relative changes in energy and angular momentum.
Differences in orbital eccentricity ($\D e$) and mean 
longitude ($\D \lam$) are for the Earth-Moon barycenter.
$\tauD$ is the divergence time (see text).
\label{fig:zb18a}
}
\end{figure*}

The astronomical solution ZB18a was originally
obtained with the integrator package \hnb\ \citep{rauch02} 
({\tt v1.0.10}) using the same setup as described above
and the symplectic integrator (2nd order WH map) with Jacobi 
coordinates \citep{zeebelourens19,zeebelourens22epsl}.
Earth's orbital eccentricity for the ZB18a solution is 
available at \myurl\ and \npurl. In order to lend
confidence to the accuracy and reproducibility of long-term 
orbital solutions for the solar system, it is imperative 
to compare the new standard solution obtained with
\orb\ (Section~\ref{sec:std}) to the original solution 
ZB18a (Fig.~\ref{fig:zb18a}). 
The changes in $|\D E/E|$ and $|\D L_z/L|$ across the 100~Myr 
integration with \orb\ and \hnb\ remain below $\sm 1\e{-11}$
and $\sm 3\e{-12}$ throughout the integrations. 

The divergence time $\tauD$ for Earth's orbital eccentricity
is \sm77~Ma (Fig.~\ref{fig:zb18a}c), again far beyond the 
reliability limit of \sm58~Ma.
$\tauD \simeq 77$~Ma even exceeds the \sm72~Ma calculated
for in- and excluding Kahan summation and symplectic correctors in 
a single integrator (\orb, Fig.~\ref{fig:std}), 
suggesting that within the limits of the current physical model
of the solar system, the performance of the integrators \orb\ 
and \hnb\ is very similar.

\subsection{Implementation of general relativity \label{sec:gr}}

As mentioned above, \orb\ includes post-Newtonian corrections 
from general relativity due to \MN\ (to 1PN order), implemented 
following ST94 (see Section~\ref{sec:pn}), which also applies to \hnb. 
However, other methods of implementing 1PN corrections are 
possible. For example, the integrator package \rebx\ provides
1PN-implementation options such as ``\grpot''
\citep{nobilirox86} and ``\gr'', based on a 1st-order 
splitting \citep[][]{tamayo20}. The option \grpot\ represents
a simplified $1/r^2$ perturbing potential, supposed
to mimic the secular advance of perihelia from GR, 
as proposed by \citet{nobilirox86}, who considered one 
out of three GR terms and ignored the other two GR terms 
that are small for the outer planets. The $1/r^2$
potential is known to incorrectly predict, for instance, 
the instantaneous elements. The option \gr\ is more 
accurate but computationally expensive (see below).
It turned out that ST94's method is either more 
accurate or significantly faster than the \rebx\ GR
options. This is not 
a criticism of the \rebx\ package, or the WH map 
as implemented in the \reb\ package \citep{rein15}. Indeed, both 
packages and their source code availability is very useful,
including for developing and testing \orbv. The 1st-order 
split
1PN-implemententation in \rebx\ had a specific intention 
\citep[for details, see][]{tamayo20} and is likely 
appropriate for many applications. However, when \gr\ is
combined with the WH map, for instance, auxiliary
computations are
required to integrate across the GR step, which results in a 
significant performance hit.
For the user with a specific 
problem at hand, it seems important to be aware of 
the differences between various options available in different 
integrator packages and their characteristics, which may
otherwise take a significant effort to figure out. 

\begin{figure*}
\begin{center}
\vspace*{-08ex}
\includegraphics[scale=0.8]{\figdir 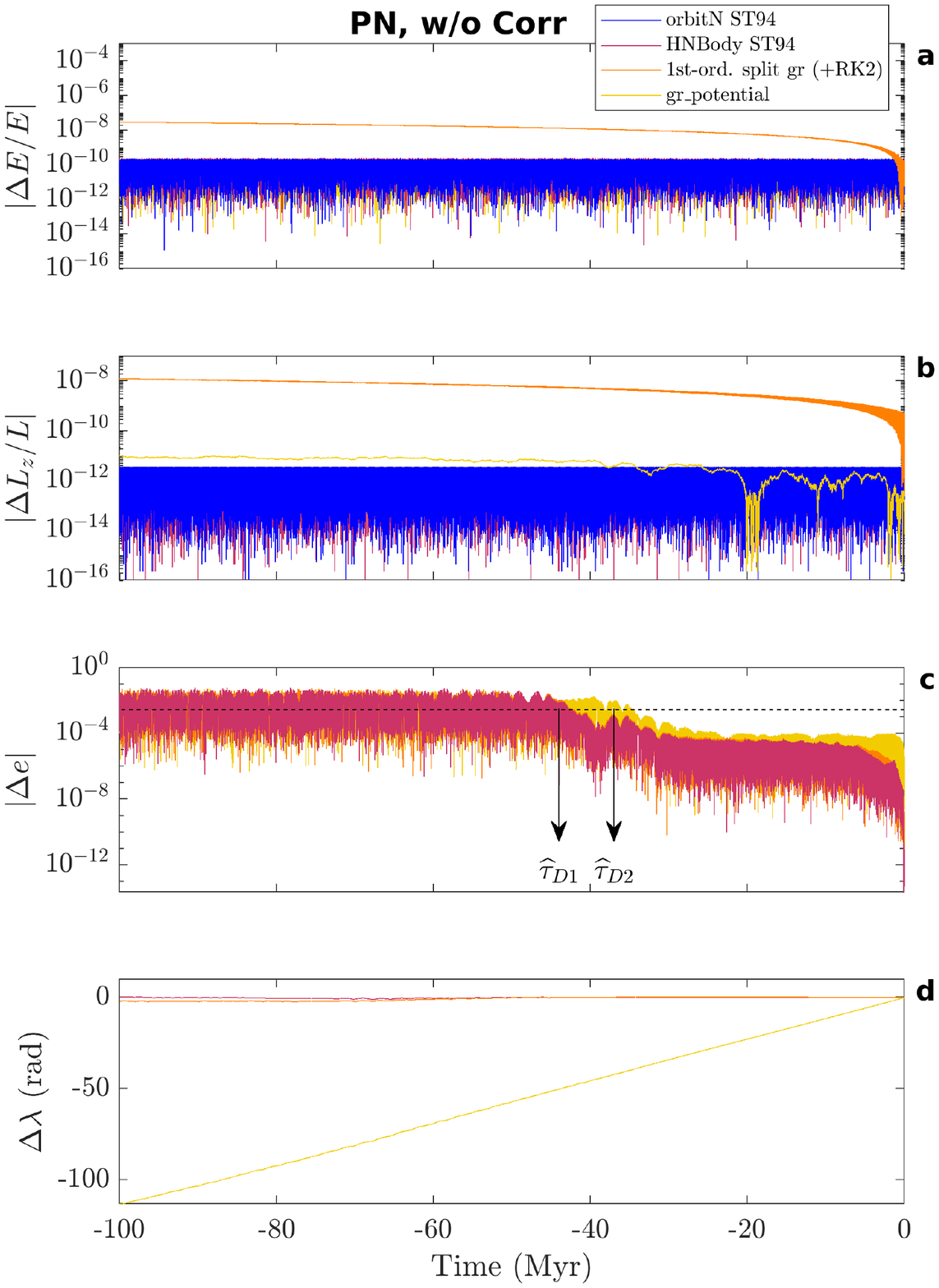}
\end{center}
\vspace*{-16ex}
%\plotone{../eps/x.pdf}
\caption{
Comparison of solar system integrations with \orb,
\hnb, and \reb\ and different 1PN implementations
(see text). The standard setup (Section~\ref{sec:std}) is 
used, except that in all packages, $J_2$, the lunar contribution 
(where available), and symplectic correctors are turned off.
$\D E/E = (E-E_0)/E_0$ and $\D L_z/L$ indicate
relative changes in energy and angular momentum.
Differences in orbital eccentricity ($\D e$) and mean 
longitude ($\D \lam$) are for the Earth-Moon barycenter.
$\tauD$'s are divergence times (see text).
\label{fig:pn}
}
\end{figure*}

Regarding
performance, it is noteworthy that the 1PN option (ST94) in 
\orb\ and \hnb\ adds less than \sm10\% computational overhead for 
the current application, while the \grpot\ and 
\gr\ ({\tt +RK2}) options add \sm4\% and \sm190\%
overhead, respectively.
Overall, for the 1PN runs displayed in Fig.~\ref{fig:pn},
\orb\ was faster than the \gr\ ({\tt +RK2}) option by a factor of \sm2.6.
Furthermore, \orb\ was about as fast (Intel i9-12900 @2.40GHz)
or \sm15\% faster (Intel i5-10600 @3.30GHz) than \hnb\ {\tt v1.0.10}, 
hence depending on hardware. All tests were performed on 64-bit Linux 
machines and {\tt gcc} optimization level 3 for \orb\ and
\reb.

To facilitate a basic comparison, the standard solar system 
integration setup (Section~\ref{sec:std})
with a few modifications was run in \orb, \hnb, and
\rebx\ (Fig.~\ref{fig:pn}). In all packages, 1PN corrections
were turned on, while $J_2$, the lunar contribution 
(where available), and symplectic correctors were turned off.
In \orb, the 1PN energy contribution was calculated according
to Eq.~(\ref{eq:hpn}), while in \rebx, the functions
{\tt rebx\_gr\_potential\_potential()} and 
{\tt rebx\_gr\_hamiltonian()}
were used. In \orb, the 1PN angular momentum was calculated following
\citet{poisson14}, while in \rebx\ no equivalent function seems
available. In \hnb, routines for 1PN energy and angular momentum
have apparently been coded (likely similar to those in \orb, 
as can be inferred from the output) but the details are
unavailable because the source code is inaccessible.

The results for $|\D E/E|$ and $|\D L_z/L|$ computed 
with \orb\ and \hnb\ (both follow ST94's 1PN implementation)
are virtually identical, while $|\D E/E|$ and $|\D L_z/L|$
(for $t \lesssim -5$~Myr) increase linearly for 
the \gr\ option (see Fig.~\ref{fig:pn}, note logarithmic 
ordinate). The differences in the 1PN implementations also
affect the orbital dynamics, showing significant differences
in the EMB's orbital eccentricity and mean longitude 
for \grpot, while the \gr\ results are closer to
those of \orb\ (Fig.~\ref{fig:pn}c and d). 
The divergence time $\tauD{}_1 \simeq 44$~Myr for the difference 
in the EMB's orbital eccentricity ($\D e$) between \orb\ and \hnb\
(Fig.~\ref{fig:pn}c) is typical for runs with 1PN corrections
enabled but $J_2$ and the lunar contribution disabled
(see Section~\ref{sec:jlp}). However, the
corresponding $\tauD{}_2 \simeq 37$~Myr between \orb\ and
\grpot\ is distinctly shorter. Thus, it
appears that the numerical 1PN implementation 
can enhance the apparent chaos in the system.

Then how can one distinguish between numerical and physical 
effects on the system's dynamics and apparent chaoticity? In other 
words, how can
one tell whether one solution is more accurate than another?
In the present case, the known limitations of \grpot\ suggest
that ST94's method is more accurate, as illustrated by differences
in the EMB's orbital eccentricity and mean longitude
(see Fig.~\ref{fig:pn}c and d). Yet, $|\D E/E|$ and 
$|\D L_z/L|$ do not hint at problems with the \grpot\ method,
as long as $|\D E/E|$ is calculated within the framework of 
the $1/r^2$ perturbing 
potential (see Fig.~\ref{fig:pn}a and b). Conversely, $|\D E/E|$ and 
$|\D L_z/L|$ for the \gr\ option may raise a red flag, yet
the calculated EMB's orbital eccentricity and mean longitude 
are closer to ST94's method.
Testing divergence times of integrators for the same perturbation
can also provide some insight to distinguish between methods
(see Section~\ref{sec:jlp}).
Further indications may be obtained by applying a different 
integrator algorithm to the same problem. For example, 
\citet{zeebe17aj} showed that solar system integrations with
\hnb\ but fundamentally different algorithms (non-symplectic
Bulirsch-Stoer method vs.\ the symplectic WH map, both 1PN 
enabled) virtually agree
to \sm63 Myr in the past, although this observation
says more about the basic algorithms than the 1PN implementation.
Thus, in summary, inspection of $|\D E/E|$ and $|\D L_z/L|$, 
as well as prior knowledge and additional tests can assist in finding 
suitable criteria to distinguish between numerical and physical 
effects on the system's apparent chaoticity. However, identifying 
such criteria is not straightforward in the present case and can be 
even trickier in other cases.

\subsection{$J_2$-, lunar-, and PN-Effects on Chaos \label{sec:jlp}}

In the following, I use \orb\ to provide insight into the 
effect of various physical processes on the long-term chaos in
the solar system, including general relativity (\pn), $J_2$
(\jtwo), and the lunar contribution (\luns). 
For the standard setup ZB18a 
(see Section~\ref{ZB18a}), the different physical effects
were turned on and off in various combinations and for 
each combination, ensemble runs
were performed ($K = 16$), with Earth's initial
$x$-coordinate perturbed by $\D x^k_0 = k \x 10^{-12}$~au 
($k = 0, \ldots, K-1$). Note that $\D x_0 = 10^{-12}$~au is much
smaller than the difference in $x_0$, say, between different 
ephemerides such as DE431 and INPOP13c
\citep[$\D x_0 \simeq 10^{-9}$~au,][]{zeebe17aj}. Next, 
$\tauD$ was determined for each 
ensemble member relative to the respective reference run 
($k = 0$). The purpose of the above procedure is to examine
the evolution of a small perturbation under the solar system's
chaotic dynamics and measure the exponential divergence
of trajectories (using $\tauD$) for different physical effects.
Tendentially, the smaller $\tauD$, the stronger the chaos. The 
ensembles provide some insight into the system's multitude of 
solutions for a set of initial conditions and allow identifying 
exceptional runs (see below).

% 2/1 column
\iftwo 
 \def\sc{0.65} 
 \def\hs{-20ex} 
\else 
 \def\sc{0.80} 
 \def\hs{000ex} 
\fi

\begin{figure}
\begin{center}
\vspace*{-28ex}
\hspace*{\hs}
\includegraphics[scale=\sc]{\figdir 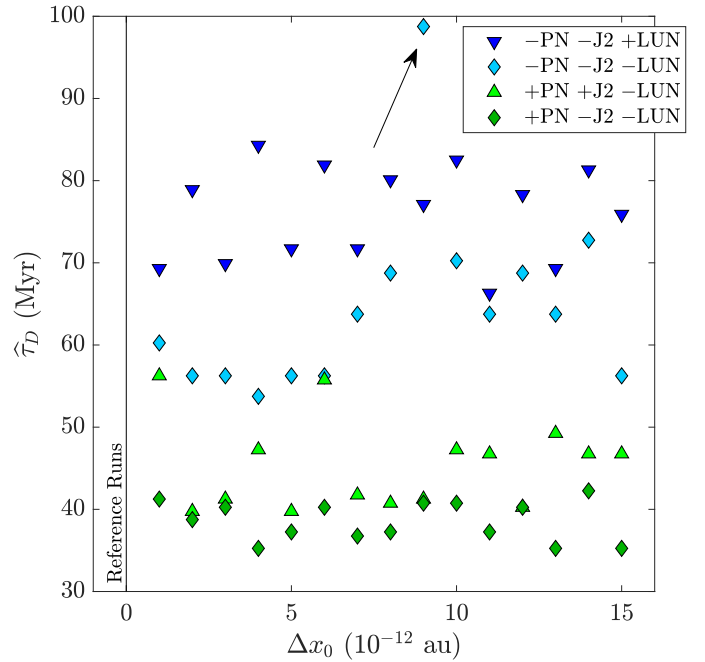}
\end{center}
\vspace*{-26ex}
%\plotone{../eps/x.pdf}
\caption{
Ensemble integrations with \orb\ turning on/off ($+/-$)
general relativity (\pn), $J_2$ (\jtwo), and the lunar 
contribution (\luns) in different combinations.
Earth's initial $x$-coordinate is perturbed by $\D x_0$,
relative to the corresponding reference run ($\D x_0 = 0$~au).
$\tauD$ is the divergence time (see text).
\label{fig:taux0}
}
\end{figure}

Turning on only \pn\ gives the smallest $\tauD$ of about
40~Myr ($+$\pn~$-$\jtwo~$-$\luns, green diamonds, 
Fig.~\ref{fig:taux0}). Adding \jtwo\ stabilizes somewhat 
(green triangles), yet removing \pn\ has an
even more stabilizing effect (blue diamonds), increasing $\tauD$
to \sm60-70~Myr ($-$\pn~$-$\jtwo~$-$\luns, Fig.~\ref{fig:taux0}).
Note that the run with $\tauD \simeq 100$~Myr ($k = 9$, 
blue diamond, arrow) is not an error or ``outlier'' that
can therefore be excluded (the run was carefully examined). 
The run is a proper solution of the system,
illustrating the inherent unpredictability of chaotic 
systems and their unaccountability in terms of conventional 
statistics. The effect of adding the lunar contribution 
($-$\pn~$-$\jtwo~$+$\luns, blue triangles, Fig.~\ref{fig:taux0}) 
increases $\tauD$ from 60-70 to $\gtrsim$ 70~Myr. The possible 
causes behind these effects, as well as the different PN 
implementations (Section~\ref{sec:gr}) in relation to the chaos 
in the system are discussed below considering the system's 
fundamental frequencies (Section~\ref{sec:frq}).

\subsection{Changes in fundamental frequencies \label{sec:frq}}

Several resonances and their overlap have been proposed 
and investigated as
the cause of the chaos in the solar system, including
the $2(g_4 - g_3) - (s_4 - s_3)$ 
resonance and the interaction between $g_5$
(largely Jupiter's forcing frequency) and $g_1$
as part of the $(g_1 - g_5) - (s_1 - s_2)$ resonance
\citep{laskar90,sussman92,morbidelli02,ito02,lithwick11,batygin15,
zeebe17aj,mogavero22,zeebe22aj,brownrein23}.
The $g$'s and $s$'s (aka fundamental, or secular frequencies,
eigenmodes, etc.) are constant in quasiperiodic systems but vary 
over time in chaotic systems, although some combinations such
as $(g_2-g_5)$ are more stable than others \citep[for discussion,
see e.g.,][]{spalding18}.
It is critical to recall that 
there is no simple 1-to-1 relationship between planet and eigenmode, 
particularly for the inner planets. The system's motion is a 
superposition of all eigenmodes, although some modes represent the 
single dominant term for some (mostly outer) planets. 
The $g_1$-$g_5$ interaction, ``$(g_1 - g_5)$'' for short,
can force Mercury's eccentricity to high values and plays
a critical role in the long-term stability of the solar 
system on Gyr time scale \citep[e.g.,][]{laskar90,batygin08,lithwick11,
zeebe15apjA,abbot23,brownrein23}. Importantly, $(g_1 - g_5)$ is 
affected by general relativity, as PN corrections move $g_1$ 
up (by \sm0.43~\asy\ at present) and away from $g_5$
\citep[for illustration, see Fig.~4 in][]{zeebe15apjA},
thus reducing the tendency for instability on Gyr time 
scale.

\begin{figure*}
\begin{center}
\vspace*{-22ex}
\includegraphics[scale=0.8]{\figdir 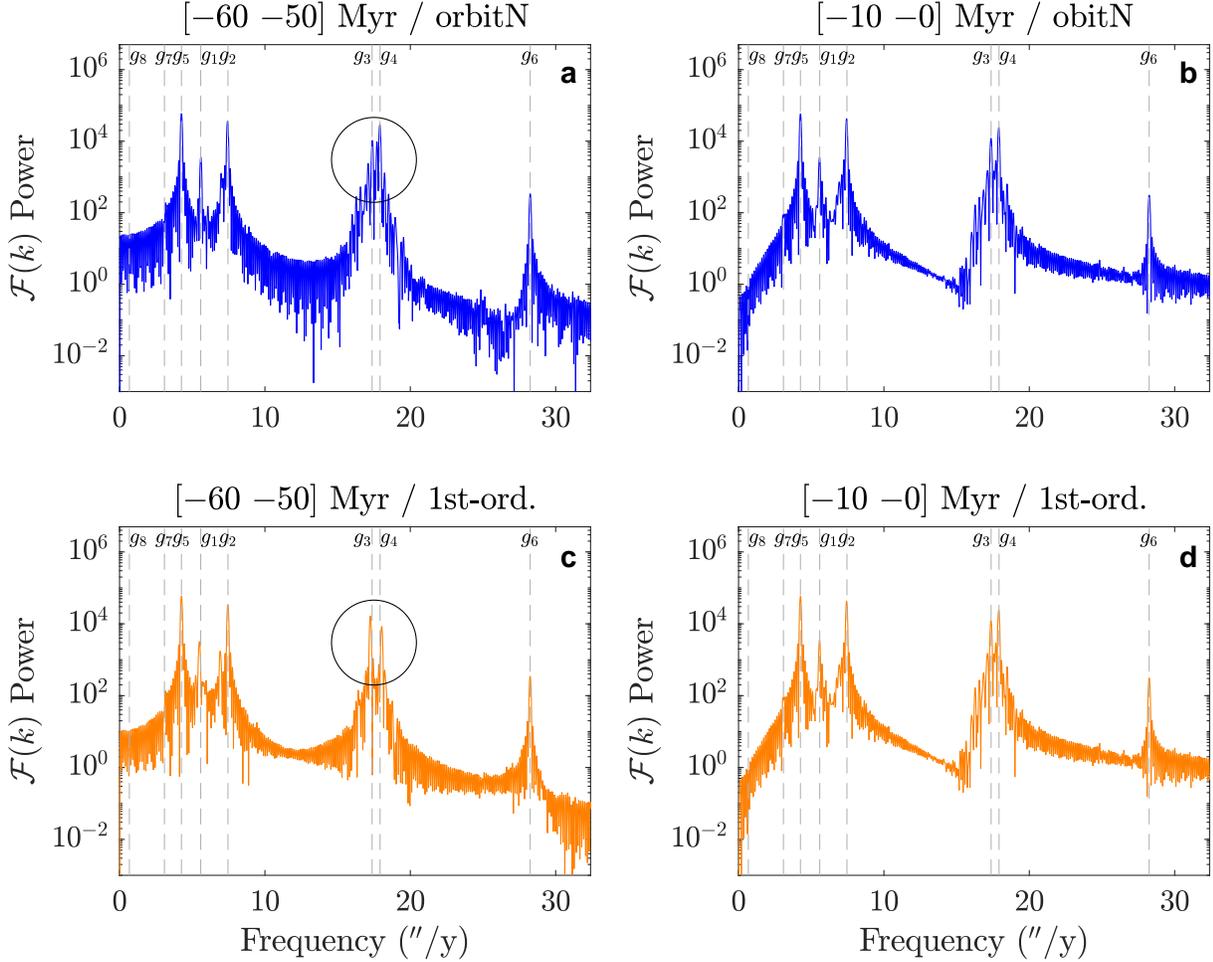}
\end{center}
\vspace*{-26ex}
%\plotone{../eps/x.pdf}
\caption{
Time series analysis of Earth's $k =  e \cos(\vpi)$ (see text) 
across the intervals $-60$ to $-50$~Myr and $-10$ to $0$~Myr
to extract solar system $g$-modes from runs with \orb\ and \reb\
\gr\ ({\tt +RK2})
for different 1PN implementations (see Fig.~\ref{fig:pn}).
$\Fc = $ Fast-Fourier Transform (FFT).
Vertical dashed lines indicate frequencies of $g$-modes 
\citep[see][]{zeebe17aj}. Note differences in $g_3$ and $g_4$
in [$-60$ $-50$]~Myr (circles in panels~a and~c).
\label{fig:fft}
}
\end{figure*}

On the contrary, PN corrections \tcr{increase chaoticity
(decrease the divergence time)}
in the current 100-Myr simulations (compare blue and green
diamonds in Fig.~\ref{fig:taux0}), suggesting that
other mechanisms, for example, the $2(g_4 - g_3) - (s_4 - s_3)$ 
resonance may be more important
on 100-Myr
time scale. Indeed, spectral analysis of, for instance, the
present integrations
comparing the 1st-order split and the symplectic 1PN implementation
(cf.\ Fig.~\ref{fig:pn}) reveal differences in $g_3$ and 
$g_4$ (see Fig.~\ref{fig:fft}), as well as small differences in 
$s_3$ and $s_4$ (not shown), consistent with a recent 
analysis of variations in Earth's and Mars' orbital inclination 
and obliquity across the same time scale \citep{zeebe22aj}.
Presuming that different resonances dominate on 
different time scales suggests a potential mechanism
for GR corrections having opposite effects on Gyr-
vs.\ 100-Myr time scale, say, through $(g_1 - g_5)$ vs.\ 
$(g_4 - g_3)$. 
As mentioned above, GR moves up $g_1$ and away from $g_5$
(the shift is much larger for $g_1$ than for $g_5$ and $g_1 > g_5$).
GR also moves up $g_3$ and $g_4$, and the shift is 
also (somewhat) larger for $g_3$ than for $g_4$ --- however, 
in this case $g_3 < g_4$. In this oversimplification,
GR would hence increase $(g_1 - g_5)$ but decrease $(g_4 - g_3)$. 
While this notion appears consistent with the results of the 
integrations performed here (summarized in Fig.~\ref{fig:taux0}),
spectral analysis of selected runs suggests a much 
more complex pattern, as detailed in the following. 
Below, spectra are presented based on the Fast-Fourier 
Transform (FFT), which was used to extract the fundamental frequencies 
($g$'s and $s$'s) from the classic variables:
\beqn
h =  e \sin(\vpi)         \quad & ; & \quad
k =  e \cos(\vpi)         \label{eqn:hk} \\
p = \sin (I/2) \ \sin \Om \quad & ; & \quad
q = \sin (I/2) \ \cos \Om \label{eqn:pq} \ ,
\eeqn
where $e$, $I$, $\vpi$, and $\Om$ are eccentricity, 
inclination, longitude of perihelion, and longitude
of ascending node, respectively.

For the two-body problem,
the change in the argument of perihelion, $\om$, due 
to GR may be written as \citep{einstein16}:
\beqn
\dot{\om} = 24\pi^3 \q{a^2}{c^2 T^2 (1-e^2)} 
             \cdot T^{-1} \ ,
\label{eq:domGR}
\eeqn
where $a$, $e$, and $T$ are the semimajor axis, eccentricity,
and orbital period; $c$ is the speed of light and the factor 
$T^{-1}$ yields $\dot{\om}$ per unit time (instead of per orbit).
Eq.~(\ref{eq:domGR}) gives $\dot{\om} \simeq 0.43$, 0.038 
and~0.014~\asy\ for 
the orbits of Mercury, Earth, and Mars at present. Spectral
analysis ($-$25 to 0~Myr) of runs with and without PN 
(blue/green diamonds,
Fig.~\ref{fig:taux0}) give a shift of $\D g_1 = 0.36$~\asy\
(Fig.~\ref{fig:dgs}a),
indicating that $g_1$ reflects Mercury's orbit but
not in a simple manner, in which case one would expect $\D g_1
= 0.43$~\asy.
For the same runs, FFT yields
$\D g_3 = 0.073$~\asy\ and
$\D g_4 = 0.055$~\asy\ (Fig.~\ref{fig:dgs}a),
showing large differences to Eq.~(\ref{eq:domGR}) (which
predicts $\dot{\om}$ in a two-body system). Thus,
while the FFT analysis shows similar tendencies to 
Eq.~(\ref{eq:domGR}), the full interacting system is much 
more complex, as expected.
As mentioned above, in general there is no simple 1-to-1 
relationship between a single planet and a single eigenmode.
Note also the large shift in $s_2$ (Fig.~\ref{fig:dgs}b).

% 2/1 column
\iftwo 
 \def\vsI{-105ex} 
 \def\vsII{-055ex} 
\else 
 \def\vsI{-088ex} 
 \def\vsII{-045ex} 
\fi

\begin{figure*}[t]
%\begin{center}
\vspace*{-40ex}
\hspace*{-08em}
\includegraphics[scale=0.7]{\figdir 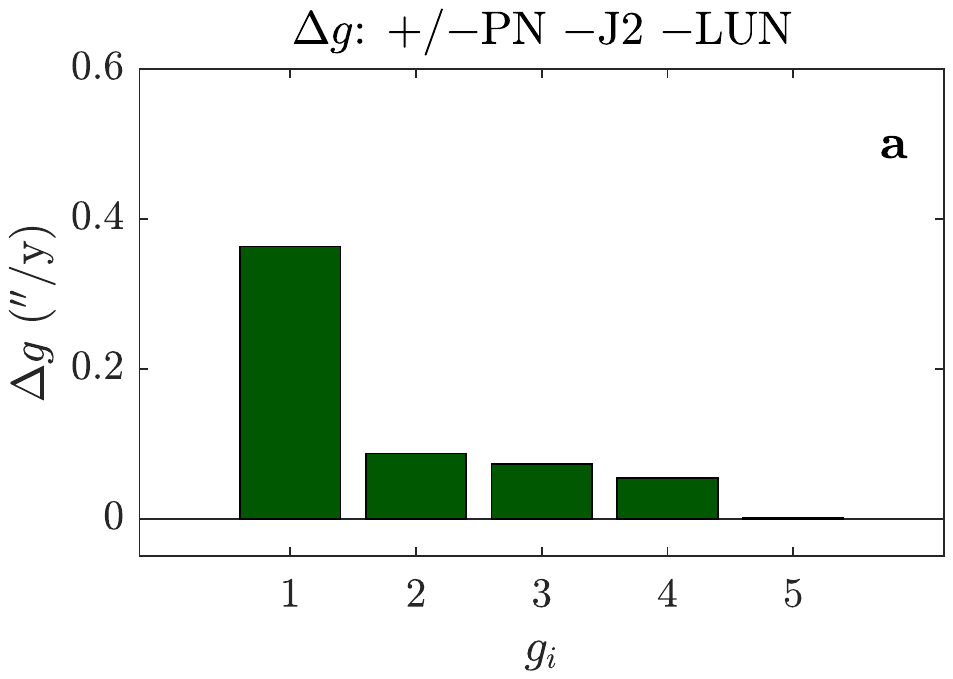}
\hspace*{-20em}
\includegraphics[scale=0.7]{\figdir 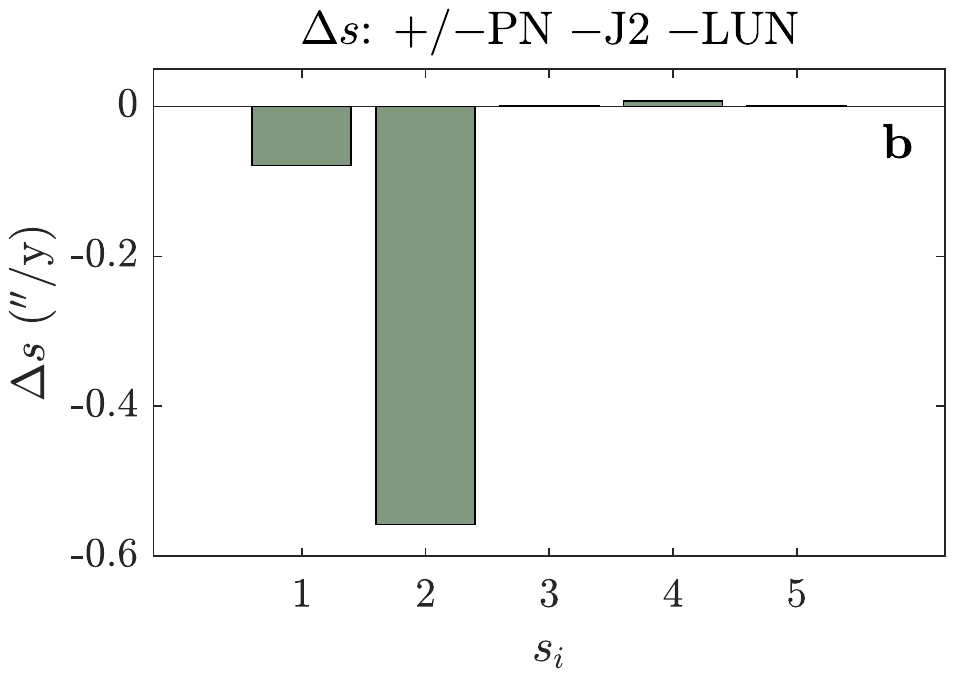}
%\end{center}

\vspace*{\vsI}
%\begin{center}
\hspace*{-08em}
\includegraphics[scale=0.7]{\figdir 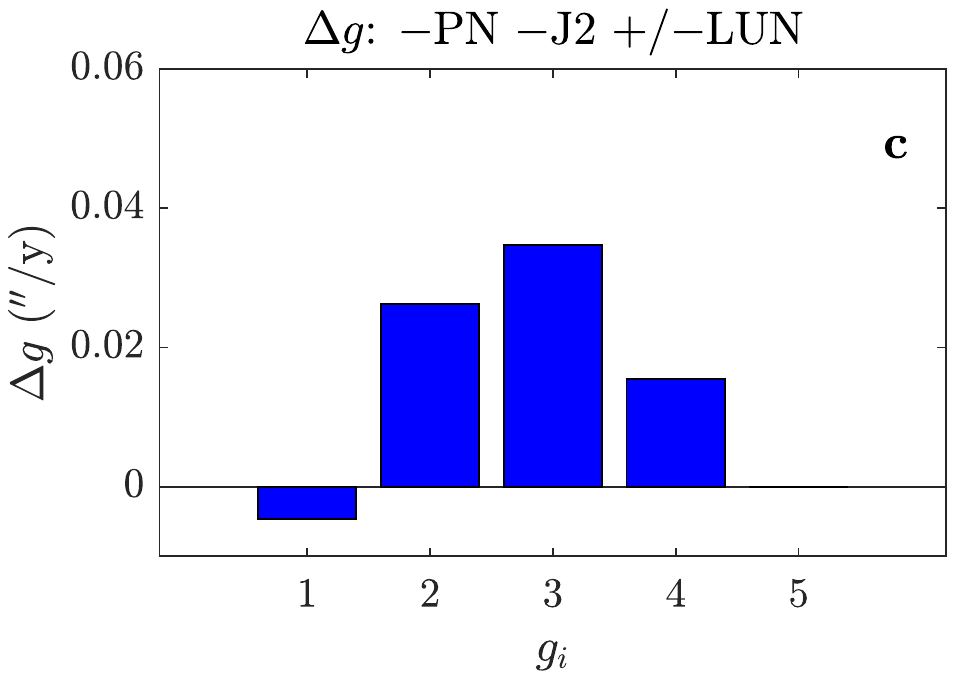}
\hspace*{-20em}
\includegraphics[scale=0.7]{\figdir 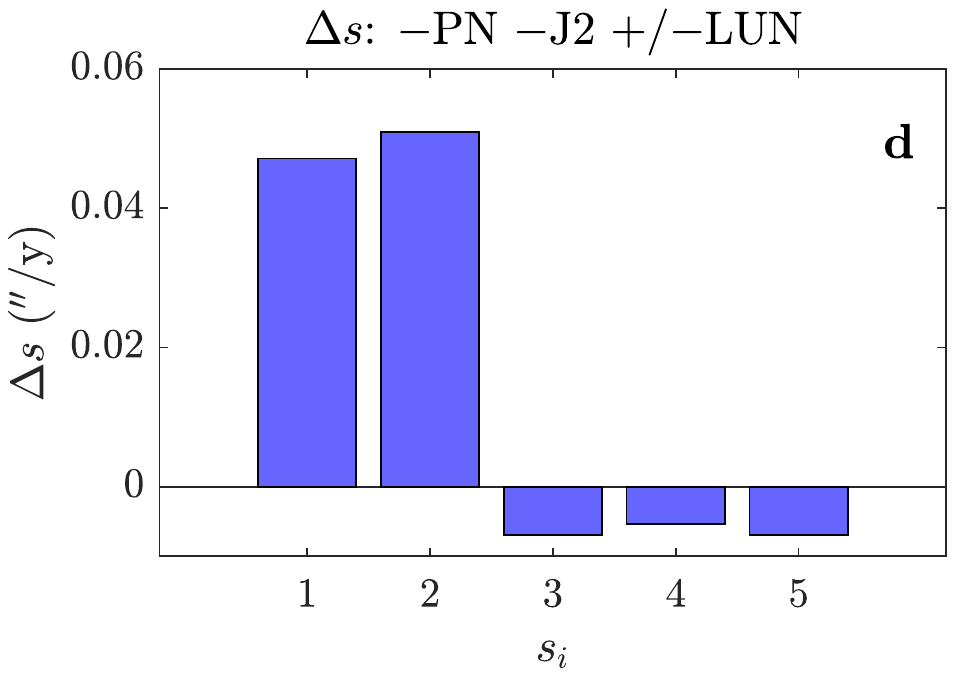}
%\end{center}

\vspace*{\vsII}
\caption{
Shifts in fundamental frequencies ($\Delta g_i$ and $\Delta s_i$)
from FFT analyses of runs shown in Fig.~\ref{fig:taux0} 
($\D x_0 = 10^{-12}$~au) across the interval 
$-$25 to 0~Myr. Shifts are calculated between two runs each with 
\pn\ on/off $ = +/-$ (green) and \luns\ on/off $ = +/-$ (blue).
\label{fig:dgs}
}
\end{figure*}

The stabilizing effect of the lunar contribution (compare 
blue diamonds and blue triangles in Fig.~\ref{fig:taux0}) 
appears similarly convoluted.
The change in the argument of perihelion of the Earth-Moon
barycenter due to the lunar contribution may be written as
(see Eq.~(\ref{eq:aql}) and Appendix~\ref{sec:omLUN}): 
\beqn
\dot{\om} = \q{n B}{a^2 (1-e^2)^2} 
\iftwo \ ; \ \else \qq ; \qq \fi
B & = & \q{3 \ m_E \ m_L \ R^2}{4 \ (m_E + m_L)^2} 
         \cdot f_L \ ,
\label{eq:domLUN}
\eeqn
where $n$ is the mean motion. Eq.~(\ref{eq:domLUN}) gives 
$\dot{\om} \simeq 0.066$~\asy\ for the EMB's orbit at present.
Spectral analysis ($-$25 to 0~Myr) of runs with and 
without \luns\ give a smaller $g_3$ shift and, in addition,
significant shifts in $g_1$, $g_2$, and $g_4$ (Fig.~\ref{fig:dgs}c),
as well as sizable shifts in $s_1$ and $s_2$ 
(Fig.~\ref{fig:dgs}d). 
Note also that the results for $\Delta g_i$
and $\Delta s_i$ as presented in Fig.~\ref{fig:dgs}
depend on the time interval selected
for spectral analysis. For example, across the interval $-$10 
to 0~Myr, $\Delta g_2$  is actually larger than $\Delta g_3$
for the \luns\ on/off case.
The numerical values of $2(g_4 - g_3) - (s_4 - s_3)$  and
$(g_1 - g_5) - (s_1 - s_2)$ are close to zero for
the runs analyzed.
In summary, it is clear that while simplified
views based on the two-body problem (e.g., Eqs.~(\ref{eq:domGR}) 
and~(\ref{eq:domLUN})) can be helpful as a starting point, they
of course fail to capture the complexity of the long-term 
dynamics of the full system.
Further in-depth analysis of the link between changes in 
fundamental frequencies, resonances, and chaos may require a 
detailed eigenmode analysis and signal reconstruction
\citep[e.g.,][]{zeebe17aj,zeebe22aj}, which is beyond the
goal of the current paper (that is, to introduce \orb) and 
is hence left for future work.

\section{Summary and Conclusions}

I have introduced the symplectic integrator 
\orb\ (version 1.0) with the primary goal of efficiently
generating accurate and reproducible long-term orbital 
solutions for near-Keplerian planetary systems dominated by a 
central mass. \orbv\ is suitable for hierarchical 
systems without close encounters but can be extended to include
additional features in future versions.
While the current \orb\ application focuses 
on the solar system, \orb\ can generally be applied to planetary systems 
with a dominant mass \MN. Among other features,
\orbv\ includes \MN's quadrupole moment, a lunar contribution, and
post-Newtonian corrections (1PN) due to \MN\ based on a fast symplectic 
implementation. I have used \orb\ to provide insight into the 
effect of various physical processes on the long-term chaos in
the solar system. The integrations performed here reveal that 
1PN corrections have the opposite effect on \tcr{chaoticity/}stability on 
100-Myr time scale, as compared to Gyr-time scale. Finally,
time series analysis was performed to examine the influence
of different physical processes on
fundamental frequencies, which affect secular resonances
and, in turn, the long-term dynamics of the solar system.

%\url{https://github.com/AASJournals/Tutorials/tree/master/Repositories}.

%\vspace*{2em}
\begin{acknowledgments}
{\bf Acknowledgments.}
I thank David Hernandez and Ilja Kocken for discussions and comments 
on an earlier version of the manuscript. I also thank the anonymous reviewer 
for suggestions, which have improved the manuscript. 
Daniel Tamayo helped clarify 1PN options in \rebx.
This research was supported by Heising-Simons Foundation Grant 
\#2021-2800 and U.S. NSF grants OCE20-01022, OCE20-34660 to R.E.Z.
\end{acknowledgments}


\software{
          \orbv\ (\giturl ; correspondence to orbitN.code@gmail.com),
          %Zenodo, doi.org/10.5281/zenodo.7933000), 
          \hnb\ {\tt v1.0.10}, \reb\ {\tt v3.19.10}, \rebx\ {\tt v3.7.1}.
          }

\iftwo
 \newpage
\fi

\appendix

\section{PN: $\alpha$-term \label{sec:alpha}}

The term $\alp_j \ \Hc^2_{Kep,j}$ in Eq.~(\ref{eq:hpnII}) leads 
to a scaling of the timestep argument of the Drift operator 
\citep{saha94} as detailed in the following. Consider 
a single body first, i.e., drop index $j$ for the time
being. Using Poisson brackets:
\beqn
\{ F , G \} = \q{\dr F}{\dr \v{x}} \q{\dr G}{\dr \v{p}}
            - \q{\dr F}{\dr \v{p}} \q{\dr G}{\dr \v{x}} \ ,
\eeqn
Hamilton's equations can be written as:
($\v{z} = [\v{x} \ \v{p}]$):
\beqn
\dot{\v{z}} =   \{ \v{z} , \Hc \} \qq \Rightarrow \qq
\dot{\v{x}} =   \q{\dr \Hc}{\dr \v{p}} \qq ; \qq
\dot{\v{p}} = - \q{\dr \Hc}{\dr \v{x}} \ .
\eeqn
If $\{ \v{z} , \Hc \} \equiv \Dc$ is considered an operator
acting on $\v{z}$, then:
\beqn
\dot{\v{z}} = \Dc \ \v{z} \ ,
\label{eq:zdot}
\eeqn
with the formal solution:
\beqn
\v{z} = \v{z}_0 \ e^{\tau \Dc} \ .
\label{eq:sln}
\eeqn
Now let $\Hc = \Hc_{Kep} + \alp \Hc^2_{Kep}$ (only affecting
the Drift operator), then
\beqn
\dot{\v{z}} \ & = & \   \{ \v{z} , \Hc_{Kep} + \alp \Hc^2_{Kep} \} \\
\dot{\v{x}} \ & = & \   \q{\dr \Hc_{Kep}}{\dr \v{p}} \ (1 + 2 \alp \Hc_{Kep} ) \\
\dot{\v{p}} \ & = & \ - \q{\dr \Hc_{Kep}}{\dr \v{q}} \ (1 + 2 \alp \Hc_{Kep} ) \ .
\eeqn
$\Hc_{Kep}$ yields the 2-body energy $-\mu m'/2a$, where $a$ is
the semimajor axis. Using $\alp = 3/(2 m' c^2)$, it follows
$2 \alp \Hc_{Kep} = - 3\mu/(2 c^2 a)$. Finally, we can write:
\beqn
\v{z} = \v{z_0} \ e^{\tau  \Dc \  (1 + 2 \alpha \Hc )}
      = \v{z_0} \ e^{\tau' \Dc} \ ,
\label{eq:slnp}
\eeqn
with $\tau' = \tau \ [1 - 3\mu/(2 c^2 a)]$. Comparing Eqs.~(\ref{eq:sln})
and~(\ref{eq:slnp}) shows that inclusion of the $\alp$-term only
changes the timestep argument of the Drift operator. Reintroducing
the body index $j$, it follows $\tau'_j = [1 - 3\mu_j/(2 c^2 a_j)]$, 
i.e., $\tau$ is scaled for each body individually, depending on mass
factor $\mu_j$ and semimajor axis $a_j$.

\section{Lunar effect on $\dot{\om}$ of EMB's orbit \label{sec:omLUN}}

The quadrupole acceleration term due to the lunar contribution
may be written as (Eq.~(\ref{eq:aql})):
\beqn
\v{a}_Q & = & - \mu_0 B \q{\v{r}}{r^5} \ ,
\eeqn
where $\mu_0 = G\MN$ and
\beqn
B & = & \q{3 \ m_E \ m_L \ R^2}{4 \ (m_E + m_L)^2} 
         \cdot f_L \ ,
\eeqn
which can be derived from a potential 
($\v{a}_Q  = -\nabla \Phi_Q$):
\beqn
\Phi_Q  & = & - \mu_0 B \q{1}{3r^3} \ .
\eeqn
Upon averaging, $r^3$ may be replaced by
$a^3 (1-e^2)^{3/2}$. Hence taking as the 
disturbing function:
\beqn
\Re  & = & \q{\mu_0 B}{3a^3 (1-e^2)^{3/2}} \ ,
\eeqn
$\dot{\om}$ is given by \citep{danby88,murraydermott99}:
\beqn
\dot{\om} = \q{n a \sqrt{1-e^2}}{\mu_0 e} \q{\dr \Re}{\dr e} \ ,
\eeqn
where $n$ is the mean motion and
$\dr \Re/ \dr i = 0$ was used ($i =$ inclination).
Finally,
\beqn
\dot{\om} = \q{n B}{a^2 (1-e^2)^2} \ .
\eeqn


%\bibliography{/home/zeebe/latex/rz.bib}{}
%\bibliography{sample631}{}
\bibliographystyle{aasjournal}

%}



\end{document}